\documentclass[12pt]{iopart}

\usepackage{bm}
\usepackage{epsfig}
\usepackage{citesort}
\usepackage{graphicx}

\newcommand{\lambdabar}%
{{\hbox{$\lambda$\kern-1.ex\raise+0.45ex\hbox{--}}}}

\long\def\dump#1{}

\begin{document}


\begin{flushright}
{\large \tt MPP-2008-20}
\end{flushright}

\title[Cosmological constraints on neutrino plus axion hot dark
  matter]{Cosmological constraints on neutrino plus axion hot dark
  matter: Update after WMAP-5}

\author{S.~Hannestad$^1$, A.~Mirizzi$^2$,
G.~G.~Raffelt$^2$ and Y.~Y.~Y.~Wong$^2$}

\address{$^1$~Department of Physics and Astronomy\\
 University of Aarhus, DK-8000 Aarhus C, Denmark\\
 $^2$~Max-Planck-Institut f\"ur Physik (Werner-Heisenberg-Institut)\\
 F\"ohringer Ring 6, D-80805 M\"unchen, Germany}

\ead{\mailto{sth@phys.au.dk}, \mailto{amirizzi@mppmu.mpg.de},
     \mailto{raffelt@mppmu.mpg.de} and \\
     \mailto{ywong@mppmu.mpg.de}}

\begin{abstract}
We update our previous constraints on two-component hot dark matter
(axions and neutrinos), including the recent WMAP 5-year data release.
Marginalising over $\sum m_\nu$ provides $m_a< 1.02 $~eV (95\% C.L.)
for the axion mass.  In the absence of axions we find $\sum m_\nu<
0.63 $~eV (95\% C.L.).
\end{abstract}

\maketitle

\noindent Very recently the Wilkinson Microwave Anisotropy Probe (WMAP) 
experiment has published its 5-year result for the angular power spectra 
of the temperature and the polarisation variations of the cosmic 
microwave background radiation, superseding their previous 3-year 
measurements~\cite{Hinshaw:2008kr,Nolta:2008ih}.  We use this 
opportunity to update our previous limits on the cosmological abundance 
of hot dark matter in the form of ordinary neutrinos as well as 
axions~\cite{Hannestad:2007dd}. We recall that axions in the eV mass 
range thermally decouple after the QCD phase transition and thus form a 
hot 
dark matter component. Cold dark matter axions, on the other hand, have 
much smaller masses,
 corresponding to a much larger Peccei--Quinn scale.  These axions are
produced nonthermally before and during the QCD phase transition. Our
update concerns the former variety of hot, thermal axions, and 
is performed for the benefit of those interested in limits
based on the latest set of data.

The analysis of the WMAP 5-year data (WMAP-5) is performed using
version~3 of the likelihood calculation package provided by the WMAP
team on the LAMBDA homepage \cite{lambda}, and follows closely the
analyses of references~\cite{Dunkley:2008ie,Komatsu:2008hk}.  Other
data sets used in this work and our parameter estimation methodology
are described in our previous paper \cite{Hannestad:2007dd}, in which
we also provide references to the literature.

\begin{figure}[b]
\hspace{25mm}
\includegraphics[width=10.cm]{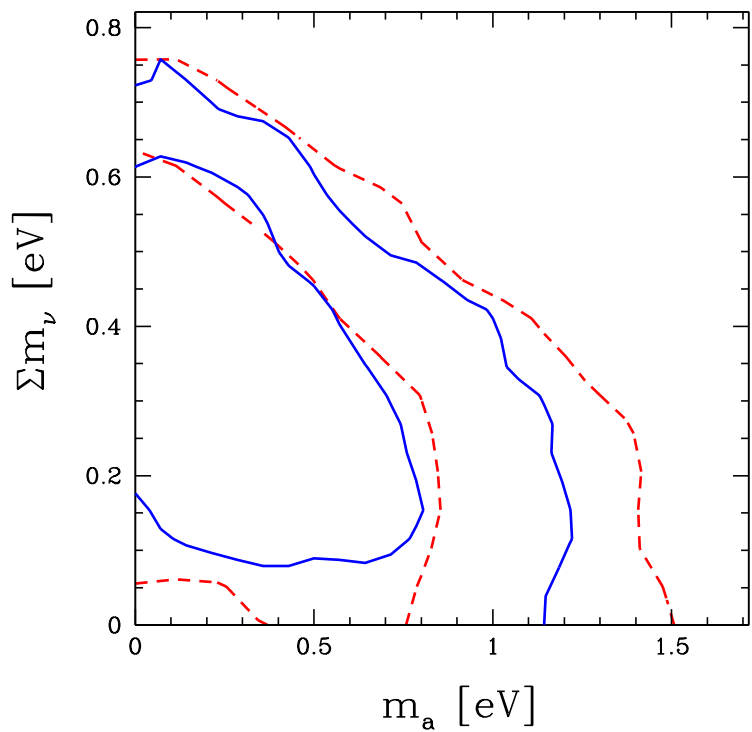}
\caption{2D marginal 68\% and 95\% contours in the $\sum m_\nu$-$m_a$
  plane derived from the standard data set of our previous paper
  (excluding Lyman-$\alpha$ forest).  The blue/solid lines correspond
  to our new results using WMAP-5, while the red/dashed lines are our
  previous constraints derived from WMAP-3.
\label{fig:contours}}
\end{figure}

We consider here only the ``baseline case'' of our previous study: We
do not include information from the Lyman-$\alpha$ forest, and use
only standard axion couplings to pions.  Figure~\ref{fig:contours}
shows our results in the form of 2D marginal 68\% and 95\% contours in
the $\sum m_\nu$-$m_a$ plane.  The blue/solid lines are our updated
contours, the red/dashed lines are our previous results using the WMAP
3-year data (WMAP-3). The modifications are very minor. Marginalising
over neutrino masses, our new axion mass limit is
\begin{equation}
m_a<1.02~{\rm eV}\quad\hbox{(95\% C.L.)},
\end{equation}
about 20\% tighter than our previous limit.  Conversely, the neutrino
mass limit after marginalisation over the axion mass is
\begin{equation}
\sum m_\nu<0.59~{\rm eV}\quad\hbox{(95\% C.L.)},
\end{equation}
which is identical to our old result.

These changes are also reflected in the 2D contours in
Fig.~\ref{fig:contours}, where a shrinkage in the allowed region in
the $m_a$ direction, but not in the $\sum m_\nu$ direction, is
apparent.  The reason for this is a better measurement of the early
integrated Sachs--Wolfe effect by WMAP-5, which in turn limits any
nonstandard effect due to extra light free-streaming species around
the epoch of matter--radiation equality.  For comparison, we also
derive a new neutrino mass limit assuming a complete absence of
axions,
\begin{equation}
\sum m_\nu<0.63~{\rm eV}\quad\hbox{(95\% C.L.)}.
\end{equation}

These limits are expected to become more restrictive with the
inclusion of smaller-scale data such as the Lyman-$\alpha$ forest
which, however, we consider too vulnerable to systematic
uncertainties. We also note that the WMAP-5 favoured value of
$\sigma_8$ is slightly larger than the WMAP-3 value
\cite{Komatsu:2008hk}. The very strong neutrino mass bound obtained
in reference \cite{Seljak:2006bg} from the inclusion of
Lyman-$\alpha$ arose partially because the $\sigma_8$ value inferred
from the SDSS Lyman-$\alpha$ data is significantly higher than that
from WMAP-3. Therefore the fact that the WMAP-5 and the SDSS
Lyman-$\alpha$ measurements of $\sigma_8$ are in better agreement
presumably will relax the hot dark matter mass bound from
CMB+Lyman-$\alpha$ data to some extent.

\section*{Acknowledgements}

We acknowledge use of computing resources from the Danish Center for
Scientific Computing (DCSC) and partial support by the European
Union under the ILIAS project (contract No.\ RII3-CT-2004-506222),
by the Deutsche Forschungsgemeinschaft under the grant TR-27
``Neutrinos and Beyond'' and by The Cluster of Excellence for
Fundamental Physics ``Origin and Structure of the Universe''
(Garching and Munich). A.M.\ is supported by a grant of the
Alexander von Humboldt Foundation.

\section*{References}

\end{document}